\documentclass[twocolumn,showpacs,unsortedaddress,aps,prl]{revtex4}
\usepackage{graphicx}

\begin{document}

\title{Thermodynamic and Transport Measurements on Superconducting
Na$_x$CoO$_2 \cdot y$H$_2$O Single Crystals Prepared by
Electrochemical De-intercalation}

\author{F.C. Chou,$^1$ J.H. Cho,$^{2\dagger}$ P.A. Lee,$^{1,2}$ E.T.
Abel,$^{1,2}$, K. Matan,$^{1,2}$ and Y.S. Lee$^{1,2}$}

\address{$^1$Center for Materials Science and Engineering, Massachusetts
Institute of Technology, Cambridge, Massachusetts 02139,\\
$^2$Department of Physics, Massachusetts Institute of Technology,
Cambridge, Massachusetts 02139}

\date{\today}

\begin{abstract}
Superconducting single crystal samples of Na$_x$CoO$_2\cdot
y$H$_2$O have been produced using an electrochemical technique
which dispenses with the usual bromine chemical de-intercalation
step and permits much more precise control of the Na content.
After careful hydration, nearly single-phase crystals have been
obtained in which over 90\% of the sample's volume corresponds to
the superconducting Na$_{0.3}$CoO$_2\cdot 1.3$H$_2$O structure.
Susceptibility and specific heat measurements confirm that bulk
superconductivity has been achieved.  The extracted normal state
density of states indicates Fermi-liquid behavior with strong mass
enhancement and a modest Wilson ratio.  Measurements of $H_{c2}$
for $H\parallel c$ and $H\parallel ab$ reveal significant
anisotropy. The estimated value of $H_{c2}$ for $H\parallel c$
yields a coherence length of $\sim100$ \AA, consistent with an
extremely narrow bandwidth.

\end{abstract}


\maketitle

Sodium cobalt oxide, Na$_x$CoO$_2$ ($x \approx$ 0.65 to 0.75), has
received considerable attention due to its unusual thermal
electric properties.\cite{1}  Recent studies have revealed
anomalous non-Fermi liquid behavior in transport properties which
point to the importance of strong correlations.\cite{2,3}  The
cobalt ions form a hexagonal layered structure and the formally
$1-x$ fraction is in the low spin $(S={1\over 2})$ Co$^{4+}$
state, while the $x$ fraction is in the $S=0$ Co$^{3+}$ state.
Interest in this material escalated with the discovery of
superconductivity by Takada {\it et al}.\cite{4} in
Na$_x$CoO$_2\cdot y$H$_2$O when the sodium concentration is
reduced to about 0.3 and water is intercalated between the layers.
The ability to control the sodium content is an exciting
development because, in principle, the limit of $x=0$ corresponds
to a Mott insulator on a triangular lattice with $S={1\over 2}$.
Then the hydrated compound can be viewed as electron doping of a
Mott insulator with a doping concentration $x \approx 0.3$. As
such, it is the second known example of superconductivity arising
from doping a Mott insulator after the high T$_c$ cuprates.  The
$S={1\over 2}$ antiferromagnet on a triangular lattice was in fact
the starting point of Anderson's resonating valence bond (RVB)
idea,\cite{5} and the new superconductor has been interpreted in
this light.\cite{6,7,8}

Until recently, the bulk properties of superconducting
Na$_{0.3}$CoO$_2\cdot 1.3$H$_2$O has been studied mostly using
powder samples.\cite{Foo,Cao,Ueland,Yang,Jin}  Single crystal
measurements \cite{Jin} are much less plentiful, though such
measurements are extremely important in order to understand this
anisotropic layered compound.  In this paper we report a new
electrochemical method to extract Na from Na$_x$CoO$_2$. This is
an alternative to the chemical de-intercalation of Na using Br
ions introduced by Takada {\it et al}.\cite{4}  Our
electrochemical method permits precise control of the Na content
and avoids the environmental hazards associated with the use of
high molar concentrations of Br.  In addition, we have succeeded
in growing large single crystals of Na$_x$CoO$_2$ by the floating
zone (FZ) method.  By applying the electrochemical procedure to
the FZ crystals, we have obtained superconducting single crystal
samples with which we have carried out a variety of physical
property measurements.

The initial Na$_{0.75}$CoO$_2$ polycrystalline material was
prepared using Na$_2$CO$_3$ and Co$_3$O$_4$ with Na to Co ratio at
0.75 to 1. The thoroughly mixed and ground powder was calcined at
750C for 12 hours and then reacted at 850C for 24 hours with
frequent grindings in between.  Na loss was minimized with a
fast-heating method.\cite{Motohashi}  A stoichiometric
Na$_{0.75}$CoO$_2$ feed rod was melted and re-crystallized with an
optical floating-zone furnace (4-Xe lamp design, CSI Japan) under
oxygen atmosphere.  Although Na vapor loss was observed from the
white deposit on the walls of the quartz sample chamber, this loss
was minimized by choosing a fast pulling rate.  A stable molten
zone can be maintained with various pulling rates from 1.5 to 10
mm/hr. Electron probe microscopy analysis (EPMA) indicates the FZ
crystal has Na content between 0.71 and 0.74 for a crystal pulled
with 2 mm/hr rate.  We have grown large single crystals of
Na$_{0.75}$CoO$_2$ (5mm diameter $\times$ 8 cm long) successfully
with this FZ method.  Powder neutron diffraction confirms that the
Na level of the FZ crystal is essentially identical to that in the
feed rod.\cite{Huang}

A three-electrode electrochemical cell was set up using the
Na$_x$CoO$_2$ sample as a working electrode, platinum foil as a
counter electrode, Ag/AgCl as a reference electrode ($E_0 =
+0.222$ V vs.~the standard hydrogen electrode), and 1M NaOH as an
electrolyte. The proposed half reaction at the anodically
polarized Na$_x$CoO$_2$ electrode responsible for Na extraction is
\begin{displaymath}
Na_xCoO_2 + \delta(OH)^- \rightarrow Na_{x-\delta}CoO_2 +
\delta(NaOH) + \delta e^-,
\end{displaymath}
where the generated NaOH becomes dissolved in the electrolyte.
Because Na$_x$CoO$_2$ is metallic, the working electrode can be
prepared directly using a compressed powder or a single crystal
sample.  Hydrated Na$_{0.3}$CoO$_2\cdot y$H$_2$O can be prepared
by having the sample anodically polarized with a constant voltage
(0.6--1.2 V) until the decaying anodic current reaches a constant
value. Alternatively, a constant current (of 0.1--10 mA) can be
used to achieve the same charge level by waiting until the final
open circuit potential reaches $\sim 0.5$ V vs.~Ag/AgCl. Samples
obtained directly from the cell are typically in a mixture of the
partially hydrated~\cite{Foo} ($c\simeq13.8$~\AA) and fully
hydrated ($c\simeq19.7$~\AA) structures and show only trace
amounts of superconductivity.  Fully hydrated superconducting
crystals can be achieved by sealing the sample in a water vapor
saturated container at room temperature.  We find that the rate of
hydration depends on grain size and several months are required to
fully hydrate mm-sized single crystals.  The sodium content in
each sample was checked by electron probe microanalysis.  After
electrochemical treatment, our as-prepared floating-zone single
crystals with $x=0.75$ typically break apart into smaller crystals
($\sim$5$\times$5$\times$2 mm) with $x\simeq0.28-0.32$.

%===========================================================
\begin{figure}
\includegraphics[width=3.2in]{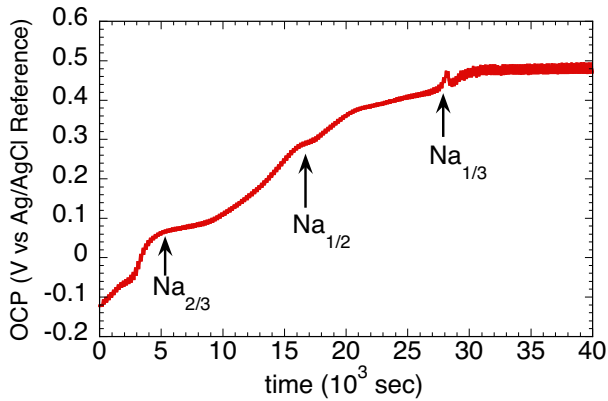}
\caption{ The quasi-open circuit potential (OCP) as a function of
time during charging of the EC cell.  The arrows indicate regions
of special stability and are labelled with possible Na
concentrations associated with these regions. } \vspace{-5mm}
\end{figure}
%===========================================================

The quasi-open circuit potential (OCP) was measured as a function
of time during a repeated sequence of having a current density of
0.01 A/g turned on (for 100 seconds) and off (for 100 seconds).
Here, the quasi OCP is the potential recorded during the time the
current is off.  We can interpret the OCP measured in this manner
as the chemical potential of the surface layer of the sample. In
Fig.~1 we plot the quasi-OCP vs.~time where the charging current
density is 0.01 A/g. The time axis is therefore proportional to
the total charge supplied to the sample surface.  A plateau in the
OCP corresponds to two-phase coexistence.  We can identify three
prominent plateaus in Fig.~1. The last plateau at 0.48 volts
indicates a saturation of the Na concentration, at least on the
surfaces of the grains of the crystal.  Assuming that this
saturation level corresponds to Na$_{1/3}$ and knowing that the
starting material is Na$_{0.75}$, we have drawn arrows in Fig.~1
to indicate the expected Na content with the assumption that the
Na extraction is proportional to the charging.  The close match of
these arrows with the plateaus suggests that stable intermediate
phases exist for Na$_{2/3}$ and Na$_{1/2}$.  We speculate that the
special fractions of $x\simeq\frac{1}{3}, \frac{1}{2},$ and
$\frac{2}{3}$ may be indicative of at least partial Na ordering
relative to the hexagonal Co structure.  The role of Na ordering
on the physics of these materials is an important subject for
future studies.

%===========================================================
\begin{figure}
\includegraphics[width=3.4in]{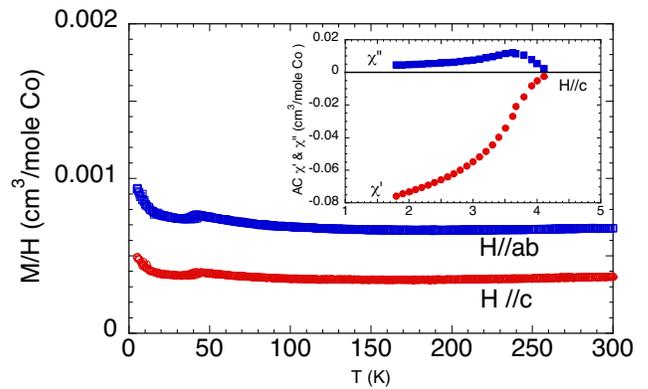}
\caption{ The magnetic susceptibility measured using a SQUID
magnetometer with an applied field of 1 Tesla with H$\parallel$ab
and H$\parallel$c.  Inset: The AC susceptibilities (in-phase
$\chi^\prime$ and out-of-phase $\chi^{\prime\prime}$) at low
temperatures.} \vspace{-6mm}
\end{figure}
%===========================================================

Magnetic susceptibility, specific heat and resistivity have been
measured on our superconducting single crystal samples.  As shown
in Fig.~2, the susceptibility shows a large anisotropy, similar to
that reported for Na$_{0.68}$CoO$_2$.\cite{2}  However, the
Curie-Weiss susceptibility that is so prominent there is now
absent.  The small cusp at 42~K is likely due to a Co$_3$O$_4$
impurity phase, which is known to have an antiferromagnetic
transition in the 33 to 46~K range.\cite{Roth} By comparing the
size the magnetization cusp with that for bulk Co$_3$O$_4$, we
estimate that the Co$_3$O$_4$ impurity fraction of our sample is
small, at the 1\% level.  However, we cannot detect the existence
of Co$_3$O$_4$ within our x-ray diffraction sensitivity.  Apart
from this anomaly and the low temperature Curie tail (which arises
from only 0.5\% of the Co moments), the susceptibility is nearly
temperature independent. Note the large magnitude and strong
anisotropy.  For comparison, the magnitude is about 5 times that
of lanthanum strontium cuprate (La$_{2-x}$Sr$_x$CuO$_4$) and 30
times that of a simple metal like Na. The anisotropy is probably
due to a combination of anisotropy in the $g$ factor and the van
Vleck term.  AC susceptibility measurements were taken in a field
of 3 Gauss/100 Hz as shown in the inset of Fig.~2. The diamagnetic
signal is indicative of a superconducting phase with an onset
temperature of about 4.2 K, and the screening fraction is
estimated to be about 120\% (without geometric correction).

The specific heat of an array of co-aligned single crystals
(combined mass of 4.7 mg) was measured using a Physical Property
Measurement System (Quantum Design) in applied magnetic fields
ranging from 0 T to 14 T and temperatures ranging from 0.37~K to
10~K as shown in Fig.~3. The sample had been hydrated for three
months after electrochemical de-intercalation, and x-ray
diffraction indicates that 94\% of the sample consists of the
fully hydrated superconducting structure. In zero field, a
pronounced peak is observed at 4.2~K, indicating the transition to
bulk superconductivity.  A second broad peak exists at lower
temperature around 1.5~K.  In a field of 1~T, the superconducting
anomaly at 4.2~K is strongly suppressed; in contrast, the peak at
1.5~K is slightly enhanced.  This suggests that the peak around
1.5~K is not related to a second superconducting phase with a
lower $T_c$.   Such a peak may be compatible with a model of
weakly interacting localized Co moments, analogous to the effects
seen in impurity-doped semiconductors.\cite{Lakner}  In fields
larger than ~5~T, both peaks disappear and are replaced by a broad
enhancement of $C/T$ over a wide range of temperatures. With
increasing field, the enhancement of $C/T$ shifts to higher
temperatures.

In the inset of Fig.~3, the top panel compares the suppression of
the superconducting anomaly for $H\parallel c$ and $H\parallel
ab$. In order to achieve a comparable suppression, an in-plane
field with magnitude 5 times that of the the out-of-plane field is
required.  This reflects the anisotropy of $H_{c2}$ in the
different field orientations and is consistent with our
resistivity measurements of $H_{c2}$ (discussed below).  We note
that $\Delta C/T_c$ in zero-field for our single crystal sample is
comparable to that reported by other groups in powder
samples.\cite{Ueland,Yang}  The data plotted in the main part of
Fig.~3 exhibit an upturn at the lowest measured temperatures which
increases with increasing field.  This is most likely due to a
Schottky contribution from the nuclear spins.  The bottom panel of
the inset shows $C/T$ data at high-fields in which a nuclear
Schottky contribution (6.7$\times$10$^{-6}$ $B^2/T^2$) has been
subtracted. These data demonstrate that in fields greater than
$\sim$10~T superconductivity is completely suppressed. Hence, 10~T
may be taken as an upper limit for $H_{c2}$ with $H\parallel ab$.

%===========================================================
\begin{figure}
\vspace{0mm}
\includegraphics[width=3.0in]{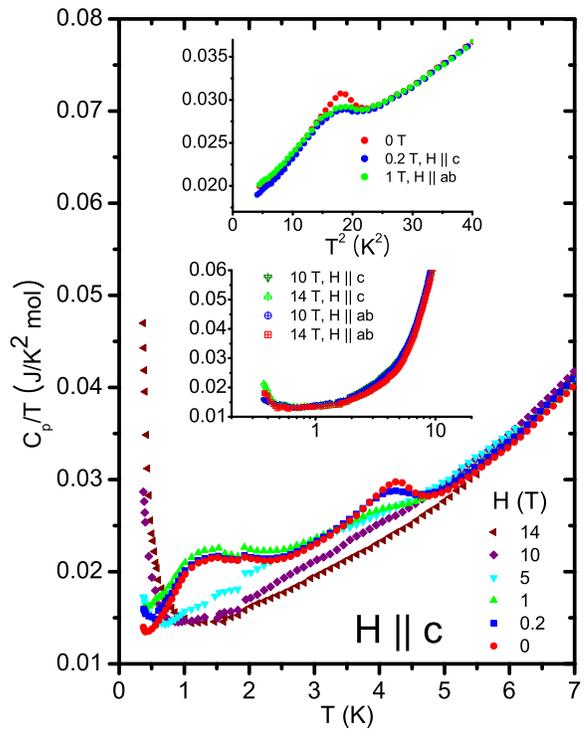}
\vspace{-2mm} \caption{ The specific heat measured on a single
crystal plotted as $c_p/T$ versus $T$ in various applied magnetic
fields. In the top panel of the inset, we plot $c_p/T$ versus
$T^2$ showing the suppression of the superconducting anomaly for
H$\parallel$c and H$\parallel$ab.  In the bottom panel of the
inset, we plot $c_p/T$ for fields of 10 T and 14 T (a nuclear
Schottky contribution has been subtracted). } \vspace{-5mm}
\end{figure}
%===========================================================

On different single crystal sample which had a smaller hydrated
phase fraction, we had observed an anomaly in the specific heat at
$T=6$~K.  This feature is also noticeable in the powder data of
Jin {\it et al}.,\cite{Jin}.  We found that this anomaly was
insensitive to magnetic fields as large as 14 T.  The origin of
this anomaly is not clear, and it may be related to another order
parameter in the vicinity of the superconducting phase on the
phase diagram.

From our specific heat data, we extract a normal state $\gamma$
value of 16.6 mJ/K$^2$ Co-mole.  This corresponds to a free
electron density of states (DOS) including both spins of 7.09
states/eV. At first sight, this seems to compare well with the LDA
band structure result\cite{14} of 4.4 states/eV.  However, this
apparent agreement is misleading because the LDA band consists of
3 overlapping $t_{2g}$ bands whereas the true quasiparticle is
expected to form a single band out of the $A_{1g}$ orbital
(symmetric combination of $t_{2g}$), split off from the rest by
correlation.  A better way is to extract a bandwidth which we
estimate to be 1.4 eV.  Even though the LDA calculation was done
for $x=0.5$ and without hydration, the bandwidth should be
insensitive to these differences.  On the other hand, we can fit
the observed DOS to that of the free electron tight binding band
on a triangular lattice with hopping matrix element $t_{eff}$.
With $t_{eff} < 0$ and $x = 0.35$, the tight binding DOS for both
spins is $0.16/|t_{eff}|$.\cite{7}  The measured DOS then implies
that $|t_{eff}| \approx$ 23 meV, or a full bandwidth of
$9|t_{eff}| \approx$ 0.2 eV, which is a factor of 7 smaller than
the LDA bandwith.  We therefore conclude that there is a mass
enhancement of $\sim 7$ compared with band theory.

By assuming $g = 2$, we find that the measured $\chi_c$ of 3.5
$\times$ 10$^{-4}$ cm$^3$/Co-mole corresponds to a free electron
DOS of 10.86 states/eV. This gives a Wilson ratio $\left(
4\pi^2k^2/3(g\mu_{B})^2 \right)\chi/\gamma$ of 1.53.  In case a
significant portion of $\chi$ comes from the van Vleck term, the
Wilson ratio will be even smaller.  The combination of
susceptibility and specific heat measurements indicate that
Na$_{0.3}$CoO$_2\cdot$yH$_2$O may be viewed as a Fermi liquid with
strong mass enhancement due to correlations. The occurrence of
superconductivity in such a narrow band material suggests an
electronic rather than phononic mechanism.

%===========================================================
\begin{figure}[t]
\includegraphics[width=3.3in]{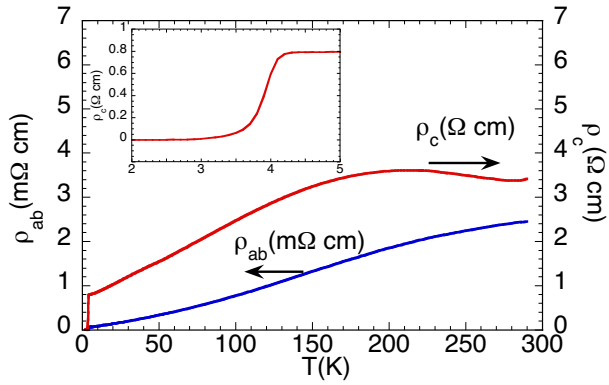}
\caption{Resistivity in the $ab$ plane and along the $c$ axis. Low
temperature results for $\rho_c$ are shown in an expanded scale in
the inset. } \vspace{-6mm}
\end{figure}
%===========================================================

Four terminal resistivity measurements on a single crystal are
shown in Fig.~4.  The zero resistance state is achieved in
$\rho_c$ and $\rho_{ab}$ at low temperatures below
$T_c\simeq4.2$~K (the inset shows an expanded view of $\rho_c(T)$
near the transition).  The large room- to low-temperature
resistance ratio ($\sim$40) indicates that the sample is a good
metal with large anisotropy between $\rho_c$ and $\rho_{ab}$
(growing to $10^4$ at low temperatures).  The conductance per
square at low T of each layer is 50 ($e^2/h$), indicative of a
long mean free path.  The current and voltage leads were attached
with silver paste which was allowed to dry in atmosphere for 12
hours. Even though partial dehydration may have occurred during
this process, it appears that superconductivity in the single
crystal sample, with its smaller surface-to-volume ratio, is more
robust than in the powder samples where superconductivity is found
to degrade in a matter of minutes.\cite{Foo}   The large peak
below 52 K reported by Jin {\it et al.}\cite{Jin} is absent.

In order to make an estimate of $H_{c2}$, we have measured the
in-plane resistance $R_{ab}$ versus $H$ at various temperatures as
shown in Fig.~5(a).  There are two regimes (at low fields and high
fields) where the resistance is roughly proportional to $H$.  We
define $H_{c2}$ as the field at the crossing point of the
extrapolation of these regimes.  In Fig.~5(b), we plot
\mbox{$H_{c2}$ vs. $T$} for both field orientations.  There is
about a factor of 5 difference in the slopes near $T_c$,
consistent with the heat capacity data.  The value of $H_{c2}$ for
$H\parallel ab$ is consistent with the Pauli paramagnetic limit
for pair breaking. From the \mbox{$H_{c2}$ vs. $T$} curve for
$H\parallel c$, we determine a coherence length of $\xi \approx
100$ \AA. This relatively short coherence length is surprising for
a superconductor with such a low T$_c$, but is entirely consistent
with the narrow bandwidth. Assuming a parabolic band, the BCS
formula for $\xi_o$ can be expressed in terms of the DOS
$\rho(\epsilon_F)$ in the following way:~\mbox{$\xi_o = {v_F\over
\pi\Delta_o} = {\sqrt{3} k_Fa^2 \over 2\pi^2
\rho(\epsilon_F)\Delta_o} \,\,.$} Using $\rho(\epsilon_F)$ = 7.09
eV$^{-1}$ extracted from specific heat and assuming $2\Delta_o =
3.52$ kT$_c$, we obtain $\xi_o \approx 27 a \approx 76$ \AA, in
reasonable agreement with the measured value.

%===========================================================
\begin{figure} [t]
\vspace{2mm}
\includegraphics[width=3.0in]{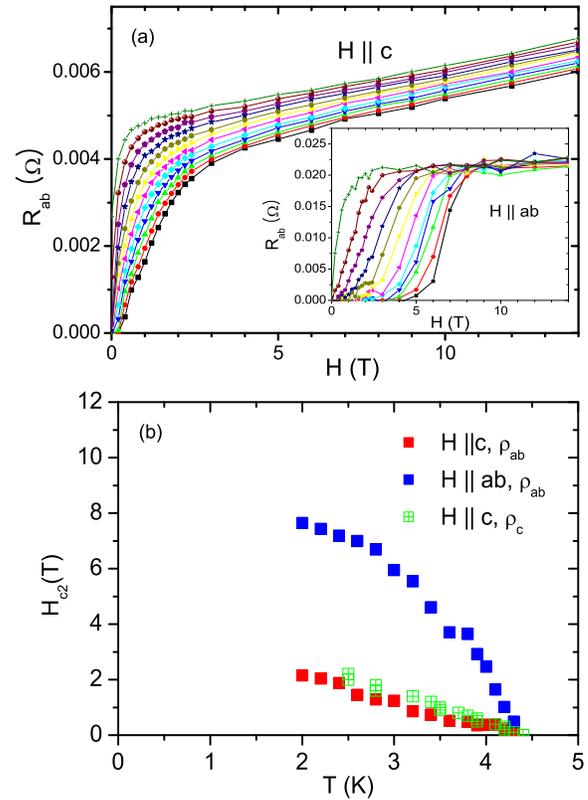}
\vspace{-3mm} \caption{(a) The in-plane resistance curves
$R_{ab}(H)$ measured for temperatures ranging from 2.0~K to 4.2~K.
Field orientations with $H\parallel c$ and $H\parallel ab$ (inset)
are measured. (b) $H_{c2}$ vs $T$ for both field orientations. }
\vspace{-6mm}
\end{figure}
%===========================================================

In conclusion, by combining a novel electrochemical method and
floating-zone crystal growth we have succeeded in producing high
quality single crystals of the hydrated Na$_x$CoO$_2 \cdot
y$H$_2$O system which show bulk superconductivity. Our
measurements indicate that the low temperature properties are
consistent with those of a Fermi liquid with strong mass
enhancement.  The availability of high quality single crystals
opens the door to many microscopic probes (such as x-ray and
neutron scattering studies) which should help achieve an
understanding of this strongly correlated material.

We thank B.~Khaykovich, R. Ott, and R. Lang for assistance with
the experimental measurements.  This research was supported by the
National Science Foundation under its MRSEC Program Award No.
02-13282 and also by Grant No. DMR 0239377.  J.H.C. was partially
supported by Korea Research Foundation Grant.(KRF-2002-
005-C20001).

\noindent$^\dagger$Permanent address: RCDAMP and Department of
Physics, Pusan National University, Pusan 609-735, Korea

%\end{multicols}
\end{document}